\documentclass[journal]{IEEEtran}
\usepackage{mathrsfs}
\usepackage{amsfonts}
\usepackage{ifpdf}
\usepackage{cite}
\ifCLASSINFOpdf
\else \fi
\usepackage[cmex10]{amsmath}
\usepackage{array}
\usepackage{mdwmath}
\usepackage{mdwtab}
\usepackage{eqparbox}
\usepackage[tight, footnotesize]{subfigure}
\usepackage{amsmath}
\usepackage{epsfig}
\usepackage{ae}
\usepackage{amssymb}
\usepackage{times}
\usepackage{amsmath}
\usepackage{booktabs}
\usepackage{color}
\usepackage{graphicx} 
\usepackage{bbm}
\usepackage[ruled,lined,linesnumbered]{algorithm2e}

\usepackage{geometry}

\begin{document}
\title{Intelligent Scheduling and Power Control for Multimedia Transmission in 5G CoMP Systems: A Dynamic Bargaining Game}
\author{Liying Li, Chungang Yang, Mbazingwa E. Mkiramweni, and Lei Pang
\thanks{L. Li, C. Yang. M. E. Mkiramweni, and L. Pang are with the State Key Laboratory
on Integrated Services Networks, Xidian University, Xi'an, 710071 China
(emails: \{llyyying@163.com; chgyang2010@163.com; mbazingwaem@yahoo.co.uk;
panglei722@163.com\}).}
\thanks{This work is supported in part by the National Science Foundation of China (61871454);
by the Fundamental Research Funds for the Central Universities(2018); by the
CETC Key Laboratory of Data Link Technology (CLDL-20182308); by the ISN02080001
and ISN90106180001; by the 111 Project under Grant B08038; and by the National
Science Foundation of China under Grant 61671062 and Grant 91638202.} }
\maketitle

\begin{abstract}
Intelligent terminals support a large number of multimedia, such as picture, audio, video, and so on. The coexistence of various multimedia makes it necessary to provide service for different requests. In this work, we consider interference-aware coordinated multi-point (CoMP) to mitigate inter-cell interference and improve total throughput in the fifth-generation (5G) mobile networks.
To select the scheduled edge users, cluster the cooperative base stations (BSs), and determine the transmitting power, a novel dynamic bargaining approach is proposed. Based on affinity propagation, we first select the users to be scheduled and the cooperative BSs serving them respectively. Then, based on the Nash bargaining solution (NBS), we develop a power control scheme considering the transmission delay, which guarantees a generalized proportional fairness among users. Simulation results demonstrate the superiority of the user-centric scheduling and power control methods in 5G CoMP systems.
\end{abstract}

\begin{IEEEkeywords}
5G, CoMP, cooperative game, delay, NBS, power control, transmission scheduling
\end{IEEEkeywords}

\section{Introduction}

With the development of the fifth-generation (5G) Generation mobile networks, the distance between two base stations (BSs) decreases significantly, which leads to the severe interference among edge users. Currently, the deployment of BSs in 5G is as follows.
Both the standalone (SA) and non-standalone (NSA) deployment are considered in the development of 5G. The NSA deployment has three phases. First, the eNB and gNB are all connected to the evolved packet core (EPC), and the control signaling of BSs is reported to EPC through the eNB. Then, the eNB and gNB are connected to EPC, and the control signaling of BSs is reported to the next-generation core network (NGCN) through the eNB; Finally, the eNB and gNB are all connected to NGCN, and the control signaling of BSs is reported to NGCN through the gNB. The NSA deployment has option3/3A, option7/7A, and option4/4A, while the SA deployment has option2 and option5. Option2 is expected to be deployed in the end stage of 5G, and its standard version was frozen in September 2018. Each deployment approach corresponds to the development phase of 5G.

With the development of 5G, BSs are deployed more densely. The ultra-dense network has led to more severe inter-cell interference, which results to serious edge users performance degradation. There are some enlightenment in the 3rd generation partnership project Rel-15. To improve the throughput of edge users, coordinated multi-point (CoMP), a promising technology, becomes more popular \cite{S1}. This technology can utilize the interference to increase the total throughput by cooperating with adjacent BSs. From the current research results, CoMP technology can by divided into coordinate schedule / beamforming and joint processing. Joint transmission (JT) is one key technology mode of JP, which enables to transmit data to one or multiple mobile stations (MSs) by multiple cooperative base stations (CBSs), and it is our research focus.

It is crucial to design an effective transmission scheduling and power control approach for increasing the throughput of edge users\cite{S2,S3}. In this paper, we optimize the transmission scheduling by studying how to cluster the BSs and how to select the CBSs serving for users. The power control is to determine the transmitting power of each BS on the overall physical resource blocks (PRBs).

Among the existing research, several traditional static transmission scheduling schemes are studied in \cite{S4,S5}. To select the edge users, an algorithm exploiting the signal processing within each cell in the interference-limited region to increase the total throughput is proposed in \cite{S4}. To enhance the performance of MSs, authors in \cite{S5} design a low-complexity algorithm combining the specific knowledge and channel state information (CSI) at the receiver for downlink CoMP. However, in real-world systems, the static clustering cannot adapt to the real channel status. Hence, some research are aimed at designing the dynamic clustering \cite{S6,S7}. Authors in \cite{S6} consider the target of sum-rate maximization in the scenario of uplink transmission. To guarantee the proportional fairness of the throughput among all users, BSs select the MSs in a round-robin way. 
Authors in \cite{S7} propose a distributed dynamic transmission scheduling algorithm exploiting reference signal receiving power (RSRP). In addition, they also propose a power control scheme based on the water filling algorithm.

Authors in \cite{S8} analyze the dynamic power allocation for downlink CoMP-NOMA (Non-orthogonal multiple access) in multi-cell networks. BSs usually allocate a great amount of resources to the edge users in NOMA, thus ensuring the fairness among users. But this algorithm is difficult to implement due to the challenge of successive interference cancellation. Meanwhile, the user who is allocated fewer resources needs to decode the better signal, which causes the larger delay. Distributed dynamic transmission scheduling in CoMP is discussed in \cite{S7}, however, it is carried out in the framework of the non-cooperative game for power allocation. Authors in \cite{S9} propose a Nash bargaining solution (NBS) fairness scheme based on cooperative game theory in orthogonal frequency division multiple access (OFDMA). A Hungarian method is proposed to allocate the optimal bargaining pairs among users.

Most existing research is based on LTE network. With the increasing bandwidth and more intelligent services, mobile applications providing videos attract more users\cite{S10}.
The network becomes denser and the quality of service(QoS) requirements in 5G are getting higher and higher.
\cite{S11} includes a general framework for the evolution of limited-feedback CoMP systems from 4G to 5G.
If the applied information is out of data before reaching the helping eNB, the performance of users will decrease. There are a lot of research on the academic field about delay.
To decrease the startup and switching delays, authors in \cite{S12} propose a solution for low delay dynamic adaptive streaming over HTTP streaming exploiting the low delay web real-time communication protocol data channel as a transport vehicle for carrying dynamic adaptive streaming over HTTP video sessions.
To avoid the applied information outdating and improve QoS, the authors in \cite{S13} study the backhaul delay problem.
Authors in \cite{S13} propose a solution for the Gaussian broadcast channels that utilizes the latest channel information available.
Small cell BSs will be deployed very densely and play a major role in minimizing delay in 5G \cite{S14}.
In this work, we mainly focus on how to decrease the transmission delay when users need to transmit video, therefore the backhaul, queuing, and other delays will be ignored.

Based on the above observations and our previous work\cite{S15}, to increase the throughput of edge users, and consider the transmission delay combining with the throughput, a new framework for intelligent scheduling and power control of multimedia transmission is proposed in 5G CoMP system.
We select the edge users according to the RSRPs and determine the cluster for each edge user in each PRB based on affinity propagation(AP) clustering algorithm \cite{S16}.
A novel power allocation game is developed based on NBS fairness. Then, we prove that there is a unique Nash equilibrium in the cooperative game.
The simulation results show the effectiveness of the proposed user-centric algorithm. 

The contributions of this paper are summarized as follows:
\begin{itemize}
\item
\textbf{AP clustering algorithm}: We model the dynamic framework based on the AP cluster algorithm in CoMP to cluster BSs for the edge users.
With this method, you do not need to manually set the center of cluster, the size of cluster, and other members of cluster.
More importantly, the algorithm can automatically adjust the cluster results according to the channel conditions.

\item
\textbf{Power allocation with NBS}: We consider fairness among edge users. Therefore, a power allocation scheme using NBS fairness is proposed in CoMP.
Especially, the utility function takes the NBS fairness into consideration.
At the same time, we define the types of edge users according to RSRPs.
\item
\textbf{The existence of maximum for utility function}: When the power received from the interference links is greater than $(1 + \sqrt 2 )$ times the system noise power, then the utility function exists maximum value.
Specifically, if the ${\rm{M}}{{\rm{S}}_1}$ and ${\rm{M}}{{\rm{S}}_2}$ belong to the same type, the power of the cluster serving ${\rm{M}}{{\rm{S}}_1}$ and ${\rm{M}}{{\rm{S}}_2}$ takes the upper bound.
If the ${\rm{M}}{{\rm{S}}_1}$ belongs to high-type and the ${\rm{M}}{{\rm{S}}_2}$ belongs to low-type, the power of the cluster serving ${\rm{M}}{{\rm{S}}_1}$ takes the upper bound.
Power of the cluster serving the ${\rm{M}}{{\rm{S}}_2}$ takes $\frac{{{\sigma ^2}}}{{n{g_3}}}$.
If the ${\rm{M}}{{\rm{S}}_1}$ is low-type and the ${\rm{M}}{{\rm{S}}_2}$ is high-type, the power of the cluster which serves the ${\rm{M}}{{\rm{S}}_1}$ takes $\frac{{{\sigma ^2}}}{{n{g_4}}}$. Power of the cluster, which serves the ${\rm{M}}{{\rm{S}}_2}$ takes the upper bound.
\item
\textbf{Combining throughput and delay}: Edge users not only need to increase throughput, but also require shorter transmission delays.
There is almost no function to consider throughput and transmission delay simultaneously.
In this paper, to emphasize the importance of delay in 5G, we optimize the transmission delay through the utility function.
Then the simulation results prove that the proposed algorithm can reduce the transmission delay and increase throughput sharply.
\end{itemize}

The rest of this paper is structured as follows. In Section \uppercase\expandafter{\romannumeral2}, the system model combined with NBS fairness and the problem description are given. In Section \uppercase\expandafter{\romannumeral3}, we present the algorithm of transmission scheduling based on AP. In Section \uppercase\expandafter{\romannumeral4}, the formulations of power control using NBS in the framework are studied.
In Section \uppercase\expandafter{\romannumeral5}, numerical results are demonstrated. Conclusions are given in Section \uppercase\expandafter{\romannumeral6}.

\section{System Model and Problem Formulation}

\subsection{System Model}

To increase the throughput of edge users, we focus on studying the performance of edge users.
In this paper, we consider the homogeneous networks CoMP with downlink transmissions and JT in the JP scenario.
Fig. 1 shows the flow of JT: First, the primary BS shares data with the helping BSs. The primary BS is the cluster center, and the helping BSs are the other members in this cluster.
Second, the BSs transmit the CSI reference signals to the user.
Third, the user transmits the CSI feedback to primary BS.
Then, the primary BS transmits the clusters relationship to the helping BSs.
Final, the BSs in the cluster transmit data to the user served by the BSs.

\begin{figure}[t]
\centering
\includegraphics[width=0.5 \textwidth]{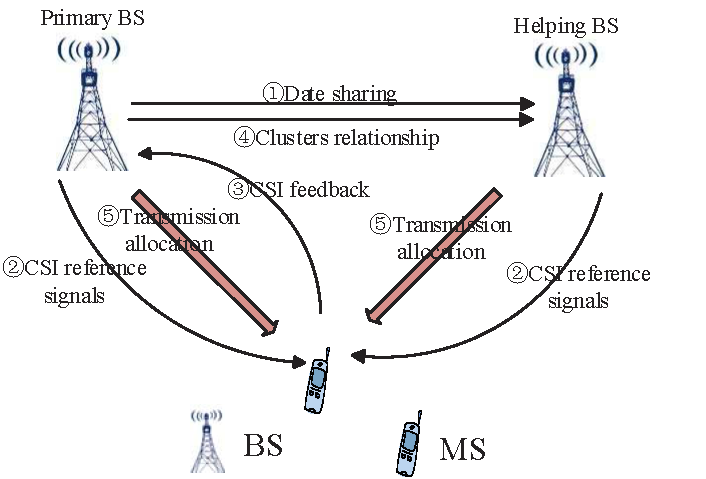}
\caption{Joint transmission in the joint processing scenario} \label{Fig.coordinate_picture2}
\end{figure}

In the case of JT, CSI is shared among all BSs simultaneously. $\textbf{\emph{N}} = \{ 1,...,N\} $ represents the set of BSs.
We consider all the MSs and the BSs work in a single antenna model, and a CBS only serves a single user.

In this work, we consider that two clusters cooperate with each other, and the BSs in the same cluster have the same transmitting power in the ultra-dense network \cite{S17,S18}.
As shown in Fig. 2a, ${\rm{M}}{{\rm{S}}_1}$ is associated with ${\rm{B}}{{\rm{S}}_2}$, and the ${\rm{M}}{{\rm{S}}_2}$ is in the coverage of ${\rm{B}}{{\rm{S}}_3}$.
The solid lines denote the interference links, and the dotted lines denote the links of the edge users receiving signal power from the associated BSs. For example, ${\rm{B}}{{\rm{S}}_{\rm{1}}} \to {\rm{M}}{{\rm{S}}_1}$ denotes the interference wireless link from ${\rm{B}}{{\rm{S}}_1}$ to ${\rm{M}}{{\rm{S}}_1}$.
The dotted lines denote the links of the edge users that receive signal power from their associated BSs. For instance, ${\rm{B}}{{\rm{S}}_{\rm{2}}} \to {\rm{M}}{{\rm{S}}_1}$ denotes the wireless link from ${\rm{B}}{{\rm{S}}_2}$ to ${\rm{M}}{{\rm{S}}_1}$.
In this scenario, ${\rm{M}}{{\rm{S}}_1}$ receives interference from ${\rm{B}}{{\rm{S}}_1}$, ${\rm{B}}{{\rm{S}}_3}$, and ${\rm{B}}{{\rm{S}}_4}$, and ${\rm{M}}{{\rm{S}}_2}$ receives the interference from ${\rm{B}}{{\rm{S}}_1}$, ${\rm{B}}{{\rm{S}}_2}$, and ${\rm{B}}{{\rm{S}}_4}$.
In Fig. 2b, the BSs are clustered into several CBSs.
The ${\rm{M}}{{\rm{S}}_1}$ is in the coverage of ${\rm{B}}{{\rm{S}}_2}$ and the ${\rm{M}}{{\rm{S}}_2}$ is in the coverage of ${\rm{B}}{{\rm{S}}_3}$. ${\rm{M}}{{\rm{S}}_1}$ receives interference from ${\rm{B}}{{\rm{S}}_3}$ and ${\rm{B}}{{\rm{S}}_4}$, ${\rm{M}}{{\rm{S}}_2}$ receives interference from ${\rm{B}}{{\rm{S}}_1}$ and ${\rm{B}}{{\rm{S}}_2}$. We define the gain of link from BS $j$ to the user $m$ as $G_{j,m}$ and the power of BS $j$ as $P_{j}$. Meanwhile, the gain from BS $j$ to the user $u$ is defined as ${G_{ju}}$.
Let the  signal to interference plus noise ratio (SINR) of ${\rm{M}}{{\rm{S}}_1}$ in the scenario of Fig. 2b be given as follows:

\begin{equation} \label{Eq:eq5}
{\rm{SINR}_{1a}} = \frac{{{P_2}{G_{21}}}}{{{P_1}{G_{11}} + {P_3}{G_{31}}{\rm{ + }}{P_4}{G_{41}}}},
\end{equation}
and the SINR of ${\rm{M}}{{\rm{S}}_1}$ in the scenario of Fig. 2b is shown as follows:

\begin{equation} \label{Eq:eq5}
{\rm{SINR}_{1b}} = \frac{{{P_1}{G_{11}} + {P_2}{G_{21}}}}{{{P_3}{G_{31}}{\rm{ + }}{P_4}{G_{41}}}},
\end{equation}

\begin{figure}[t]
\centering
\includegraphics[width=0.5 \textwidth]{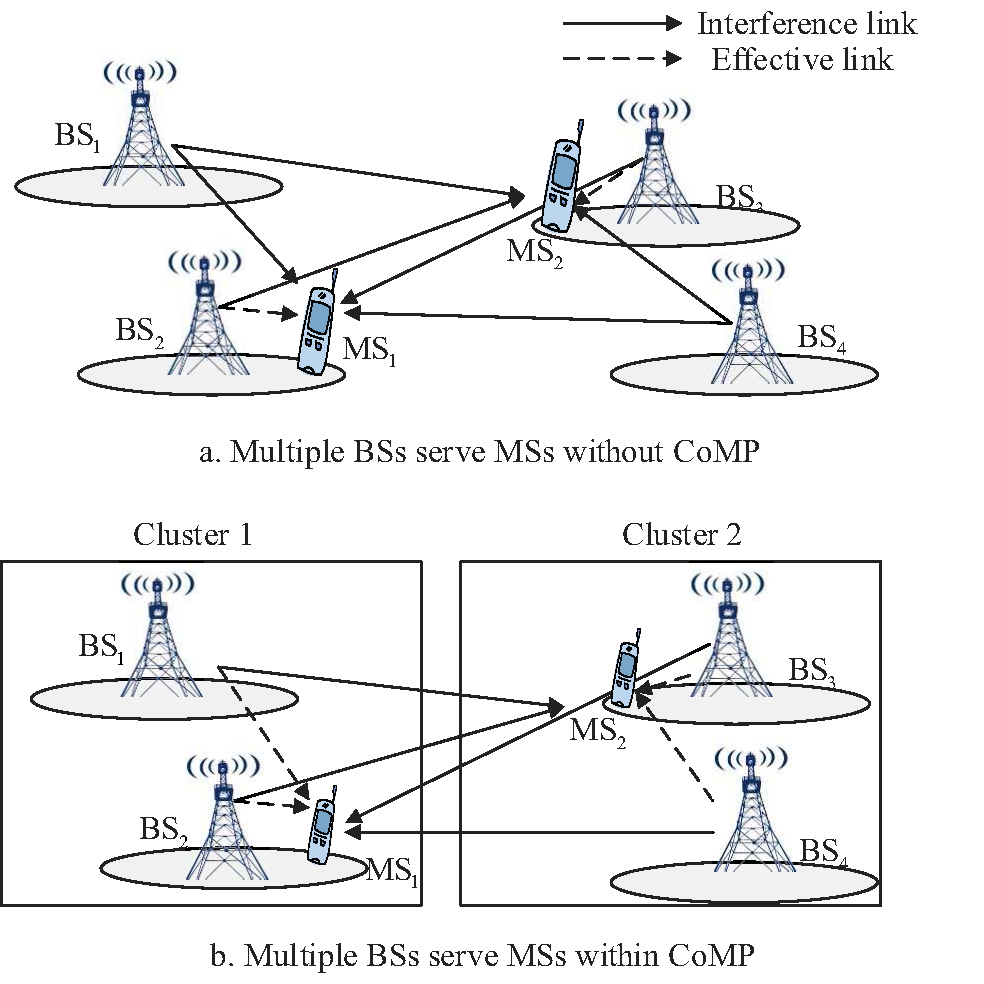}
\caption{Comparison of the interference under CoMP with the interference without CoMP} \label{Fig:fig1}
\end{figure}

By comparing (1) and (2), we find that
\begin{equation} \label{Eq:eq5}
\begin{aligned}
{\rm{SINR}_{1a}} =& \frac{{{P_2}{G_{21}}}}{{{P_1}{G_{11}} + {P_3}{G_{31}} + {P_4}{G_{41}}}} \\
<& \frac{{{P_1}{G_{11}} + {P_2}{G_{21}}}}{{{P_1}{G_{11}} + {P_3}{G_{31}} + {P_4}{G_{41}}}} \\
<& \frac{{{P_1}{G_{11}} + {P_2}{G_{21}}}}{{{P_3}{G_{31}} + {P_4}{G_{41}}}} = {\rm{SINR}_{1b}},
\end{aligned}
\end{equation}
it can be seen that ${\rm{SINR}_{1a}}<{\rm{SINR}_{1b}}$. Hence, the interference of the users with CoMP technology is lower than the users without CoMP technology.

\subsection{Problem Formulation}

CoMP requires the processing devices of multiple sites to work closely together. This is precisely the advantage of centralized radio access network (C-RAN).
C-RAN is an architecture based on centralized processing, cooperative radio, and real-time cloud infrastructure.
In addition, the architecture based on centralized unit (CU) and distributed unit (DU) has also been recognized by the industry. Therefore, the deployment with CU / DU separation is widely used in 5G deployment due to its unique advantages.
The advantages are mainly reflected in the different deployment locations of CU to allocate different resources for traffic diversity. Besides, as networks become denser, the signaling interacting cannot balance for so many points, and when there is a decision conflict, the authority of peer points is not enough. But the balance and authority problem will be solved through pooling DU.

As for dense networks, to satisfy users' high bandwidth requirements, it is necessary that there are not only macro base station (MBS) but also small base station (SBS), as depicted in Fig. 3.
Users can be served by MBS or SBS separately or simultaneously. Because SBSs can get more comprehensive coverage, so they are commonly used.

(A) no CoMP:

For BSs serve MSs under non-CoMP scenario, one MS only served by one BS. In general, center users are more suitable for this service mode as shown in Fig.3(A);

(B) CoMP with NSA:

For CoMP under NSA deployment, MBSs and SBSs serve MSs simultaneously as shown in Fig.3(B). A large amount of information needs to interact in ultra-dense networks. Therefore, the integration of CU is profitable;

(C) CoMP with SA:

For CoMP under SA deployment, with the development of 5G, the ultimate goal is to develop into SA deployment as shown in Fig.3(C).

\begin{figure}[t]
\centering
\includegraphics[width=0.4 \textwidth]{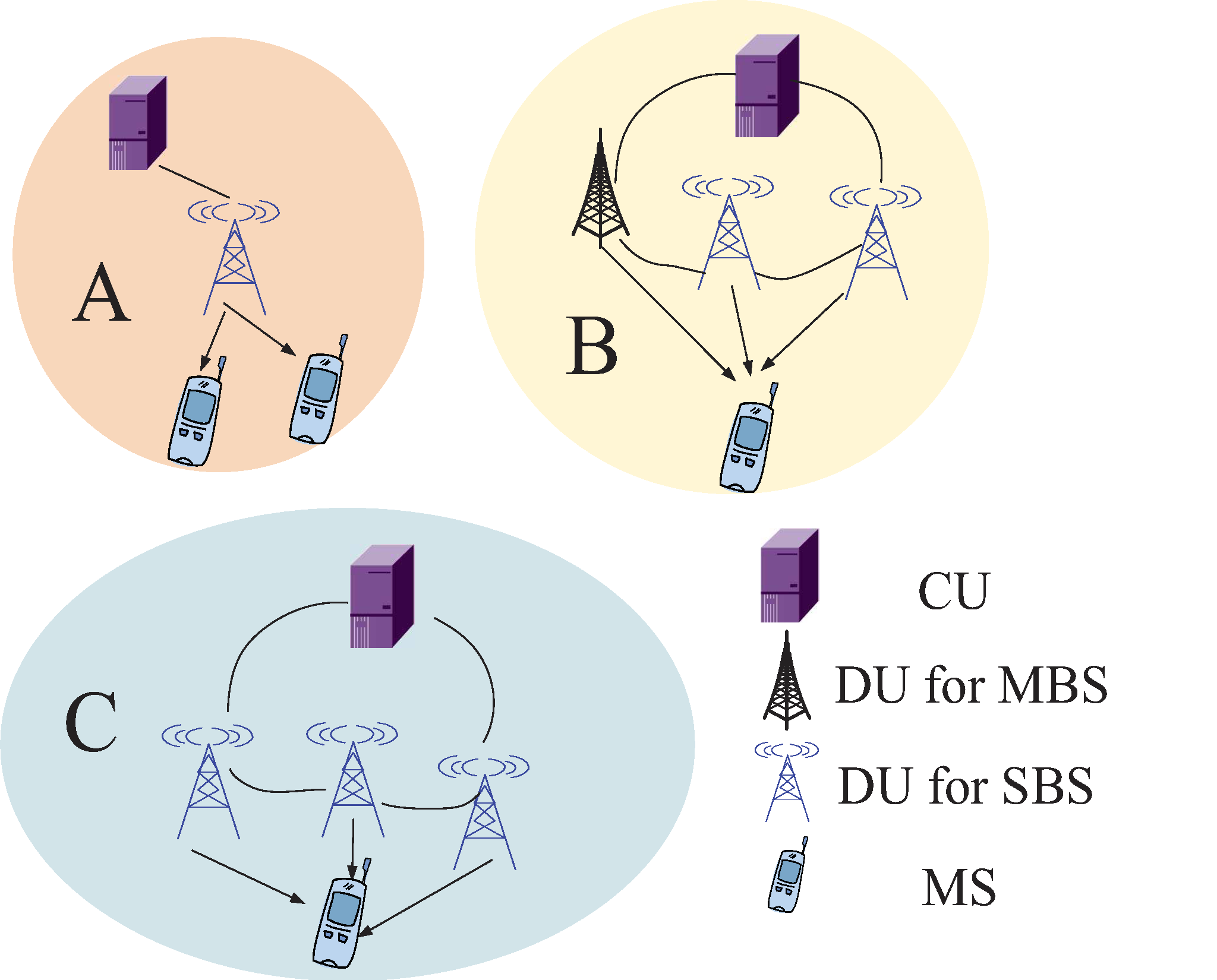}
\caption{CoMP scenario in 5G} \label{Fig:fig1}
\end{figure}

We define $\textbf{\emph{R}} = \{ 1,...,R\} $ as the set of PRBs, and $\textbf{\emph{M}}=\{1,...,M\}$ as the set of users. Let the BSs in the k${th}$ CBS be a set ${\rm CBS}_{b,k}$ for $ k_b \ge k \ge 1$, where ${k_b}$ is the total number of CBSs in the PRB $b$. We set ${P_{bj}}$ as transmission power allocation of BSs in ${\rm{CB}}{{\rm{S}}_{b,k}}$. Let the system noise power be ${\sigma ^2}$. In this scenario, ${g_{b,j,k}}$ is defined as the power gain of the link from BS $j$ to the relevant user in PRB $b$.
SINR of user scheduled in PRB $b$ and served by ${\rm{CB}}{{\rm{S}}_{b,k}}$ as ${\gamma _{b,k}}$ is presented in (4).

\begin{equation} \label{Eq:eq5}
{\gamma _{b,k}} = \frac{{\sum\limits_{j \in {\rm{CB}}{{\rm{S}}_{b,k}}} {{P_{b,j}}{g_{b,j,k}}} }}{{\sum\limits_{j^\prime \notin {\rm{CB}}{{\rm{S}}_{b,k}}} {{P_{b,j}}{g_{b,j^\prime k}}}  + {\sigma ^2}}}.
\end{equation}

We define the rate of the user served by ${\rm{CB}}{{\rm{S}}_{b,k}}$ in the PRB $b$ as ${{\rm{R}}_{b,k}}$, which is
\begin{equation} \label{Eq:eq5}
{{\rm{R}}_{b,k}} = \frac{B}{R}{\log _2}(1 + {\gamma _{b,k}}),
\end{equation}
where the system bandwidth defined as $B$, and $R$ represents the number of PRBs.

\subsection{Initial Analysis}

We focus on increasing the throughput of edge users. It can be described precisely as follows:
\begin{equation} \label{Eq:eq5}
\max \sum\limits_{b \in \textbf{\emph{R}}} {\sum\limits_{1 \le k \le {k_b}} {{\rm{R}}_{b,k}} },
\end{equation}

\begin{equation} \label{Eq:eq5}
{\rm{s}}{\rm{.t}}{\rm{.  }}\sum\limits_{b \in \textbf{\emph{R}}} {{P_{b,j}}}  \le {p_{\max }},\forall j \in \textbf{\emph{N}},
\end{equation}
where ${P_{b,j}}$ represents the power of BS $j$ on the PRB $b$, and $p_{max}$ is the maximal power constraint at each BS $j$.

Since the formulated problem is non-concave, it is hard to solve.
To solve this problem, it can be divided into two sequential sub-problems, including how to cluster and determine the transmitting power of the BSs transmitting to users in the clusters.
The former subproblem will be solved the heuristic algorithms in Section \uppercase\expandafter{\romannumeral3}, and the latter will be discussed in Section \uppercase\expandafter{\romannumeral4}.

\subsection{ NBS for the CoMP system}
We focus on the particular strategy of proportional fairness \cite{S19}, which is closely linked to the game-theoretic concept of NBS fairness. In this section, we will introduce how to use the NBS fairness in CoMP \cite{S9}. We formulate the power allocation problem using NBS as follows:

\begin{equation} \label{Eq:eq2}
\begin{array}{l}
\max \prod\limits_{i = 1}^L {({U_i} - U_i^{\min })} \\
\begin{aligned}
s.t.\quad&{\rm{C1:}}\quad {p_m} \ge 0,m = 1,2, \cdots ,L\\
&{\rm{                C2:   }}\quad {p_m}{g_n} \ge {p_0},m = 1,2, \cdots ,L , n = 1,2, \cdots,N \\
&{\rm{                C3:   }}\quad {p_m} \le {p_{\max }},m = 1,2, \cdots ,L\\
&{\rm{                C4:   }}\quad {U_i} \ge U_i^{\min },\\
\end{aligned}
\end{array}
\end{equation}
where $L$ is the number of players and the ${U_i}$ is the profit of $i$th player. Let $U_i^{\min }$ represent the minimal payoff that the $i$th player.

Constraint C1 is to ensure the transmission power is non-negative. Constraint C2 shows RSRPs that each MS receives is no less than a threshold ${p_0}$. Otherwise, the MS cannot decode the signal. Constraint C3 shows the total maximal transmitting power constraint ${p_{max}}$ at each BS. Constraint C4 guarantees the utility function is always greater than $U_i^{\min }$.

Our work is to solve the optimization problem (8), that is how to increase the throughput of edge users and decrease the transmission delay of edge users overall BSs around PRBs, and how to determine the transmitting power of each BS in the clusters.

\section{Transmission Scheduling}
In this section, we propose a user-centric dynamic transmission scheduling scheme.
\begin{figure}[t]
\centering
\includegraphics[width=0.4 \textwidth]{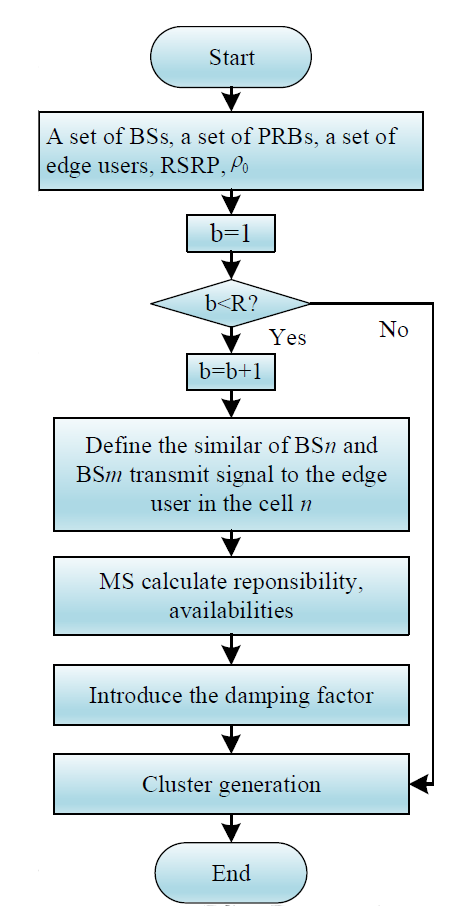}
\caption{Flowchart of AP cluster algorithm} \label{Fig:fig1}
\end{figure}

First, input a set of BSs, a set of PRBs, a set of edge users, RSRP, and ${\rho _0}$.
Second, the nodes in the AP cluster represent BSs, including MBSs or SBSs. We define the real-valued similarity of $S(m,n)$ as the BS $n$ and the BS $m$ transmit a signal to the edge user in the cell $n$.
Third, MS calculates the responsibility based on $R(m,n)=S(m,n)-\mathop {\max }\limits_{n' \ne n} \{ A(m,n') + S(m,n')\} $. We define availabilities as $A(m,n)=\min \{ 0,R(n,n) + \sum\limits_{m' \notin \{ m,n\} } {\max (0,R(m',n))} \} $ when ${m \ne n }$. Otherwise, we define $A(m,n) = \sum\limits_{m' \ne n} {\max \left( {0,R(m',n)} \right)} $.
To prevent large variations of parameters, we introduce the damping factor ${\lambda}$. In this paper, damping factor is used in $R = \lambda {R^{old}} + (1 - \lambda ){R^{new}}$ and $A = \lambda {A^{old}} + (1 - \lambda ){A^{new}}$ \cite{S20}.
Then, stop iterating when the result of the clusters is constant. At any point find the value of maximal $(A(m,n)+R(m,n))$, where the point $n$ is the cluster center of the point $m$, and the point $m$ is the cluster member of the cluster center of $n$.
Final, if RSRP of the MS received from BSs in the cluster is lower than the threshold, the BS is removed from the cluster which served the MS. Then, the center of the cluster is the edge user, and the others in the cluster are scheduled BSs which serve the edge user.
\section{Problem Formulation Based on NBS}
\subsection{ Two-User Multi-BS Case Power Control under NBS}
For this part, we consider that there are two clusters in the CoMP system, and the transmission power of each BS is equal in the same cluster. That is, the power of ${\rm{B}}{{\rm{S}}_1}$ is the same as the power of ${\rm{B}}{{\rm{S}}_2}$ in Fig. 5.
We define the transmitting power of the BSs in cluster 1 as ${p_1}$, and the transmitting power of BSs in cluster 2 as ${p_2}$.
To simplify the system model, let the channel gain is approximate consistent among the interference link, such as define the gain of ${\rm{U}}{{\rm{E}}_1}$ received the ${\rm{B}}{{\rm{S}}_2}$ covered in the cluster as ${g_{\rm{1}}}$ in Fig. 5.
We define the $i$th user minimal payoff as $T _{i \min }$.
Under the model of NBS, we consider that the utility function of the users in cluster 1 is ${T _1}$, the utility function of the users in cluster 2 is ${T _2}$.

\begin{algorithm}[t]
\caption{Two-User Case Power Control under NBS}
\textbf{Input}\\
RSRPs, ${\rho _0}$, scheduled MSs, n(number of BSs in the CBSs served MSs), the types of the scheduled MSs\\
 \textbf{Output}\\
Transmission power\\
\textbf{1. Initialization}\\
Set $N \to \textbf{\emph{B}}$ and $M \to \textbf{\emph{A}}$ \\
\textbf{2. while}
existence the maximal value\\
\textbf{do}\\
\textbf{if} the ${\rm{M}}{{\rm{S}}_1}$ and ${\rm{M}}{{\rm{S}}_1}$ are the same types.\\
Power of the cluster, which serves the ${\rm{M}}{{\rm{S}}_1}$, takes the upper bound.\\
Power of the cluster, which serves the ${\rm{M}}{{\rm{S}}_2}$, takes the upper bound.\\
\textbf{end}\\
\textbf{if} the ${\rm{M}}{{\rm{S}}_1}$ is high-type and the ${\rm{M}}{{\rm{S}}_1}$ is low-type.\\
Power of the cluster serving the ${\rm{M}}{{\rm{S}}_1}$ takes the upper bound.\\
Power of the cluster serving the ${\rm{M}}{{\rm{S}}_2}$ takes $\frac{{{\sigma ^2}}}{{n{g_3}}}$.\\
\textbf{end}\\
\textbf{if} the ${\rm{M}}{{\rm{S}}_1}$ is low-type and the ${\rm{M}}{{\rm{S}}_1}$ is high-type.\\
Power of the cluster serving the ${\rm{M}}{{\rm{S}}_1}$, takes $\frac{{{\sigma ^2}}}{{n{g_4}}}$.\\
Power of the cluster serving the ${\rm{M}}{{\rm{S}}_2}$ reaches the upper bound.\\
\textbf{end}\\
\end{algorithm}

The utility function in the model of NBS is shown as follows:
\begin{equation} \label{Eq:eq2}
\begin{array}{l}
U = {({T _1} - {T _{1 \min}})^{{w_1}}}{({T _2} - {T _{2 \min }})^{{w_2}}}\\
\begin{aligned}
s.t.\quad&{\rm{C1:}}\quad {\rm{ }}{p_m} \ge 0,m = 1,2\\
&{\rm{                C2:   }}\quad {\rm{     }}{p_m}{g_{n}} \ge {p_0},n = 1,2,3,4,m = 1,2\\
&{\rm{                C3:   }}\quad {\rm{     }}{p_m} \le {p_{\max },m = 1,2}\\
&{\rm{                C4:   }}\quad {T_i} \ge T _{i \min}
\end{aligned}
\end{array},
\end{equation}
where the $w{}_i$ is the weight of user $i$. The RSRP is defined as $\rho $. We divide these users into the high-type and low-type based on ${\rho _0}$, which depends on the threshold value of RSRP \cite{S21}.
If the user's  RSRP is higher than ${\rho _0}$, we define the user as a high-type user, otherwise we define the user as a low-type user.

 Considering that the weight of high-type user is ${w_H} = 2$, and the weight of low-type user is ${w_L} = 1$, i.e.,
\begin{equation} \label{Eq:eq6}
{w_H} - {w_L} = 1.
\end{equation}

\begin{figure}[t]
\centering
\includegraphics[width=0.45 \textwidth]{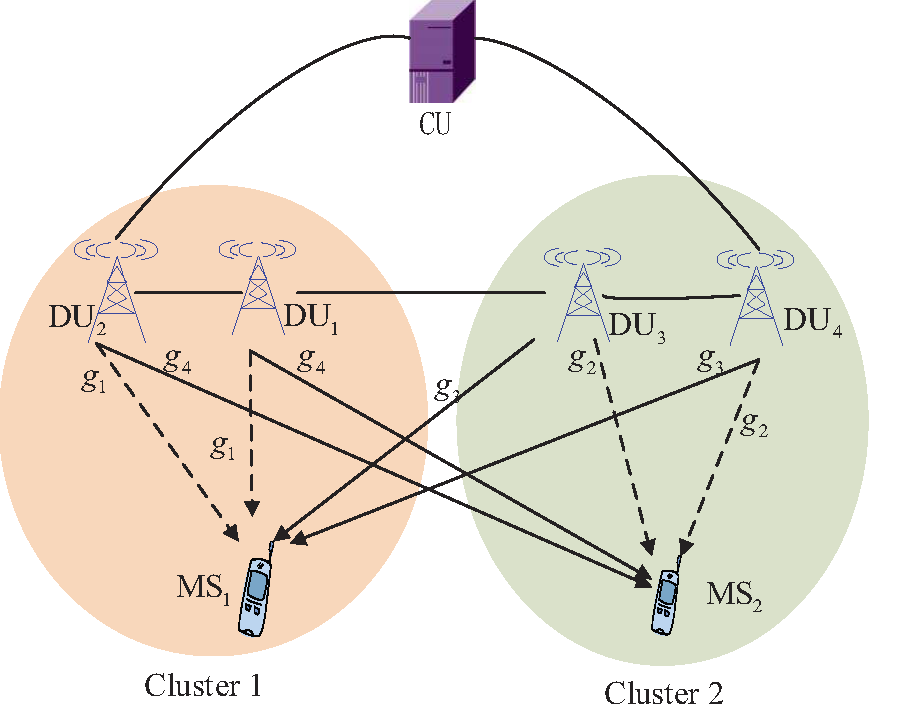}
\caption{System model for CoMP} \label{Fig:fig5}
\end{figure}
In ultra-dense networks, to guarantee the QoS of users, transmission delay plays a major role in 5G.
To take the throughput and the transmission delay into consideration simultaneously, we define the utility function as follows:
\begin{equation} \label{Eq:eq4}
\begin{aligned}
T = {\rm{SINR}} \times {{\rm{e}}^{\frac{1}{{Delay}}}}
\end{aligned},
\end{equation}
where the delay of user can be estimated through dividing the size of files by the user's transmission rate.
The utility function of user 1 and user 2 can be expressed in detail as follows:

\begin{equation} \label{Eq:eq4}
\begin{aligned}
{T_1} &= \frac{{n{p_1}{g_1}}}{{n{p_2}{g_3} + {\sigma ^2}}} \times {e^{\frac{{\frac{B}{R}\log (1 + \frac{{n{p_1}{g_1}}}{{n{p_2}{g_3} + {\sigma ^2}}})}}{M}}}\\
 &= \frac{{n{p_1}{g_1}}}{{n{p_2}{g_3} + {\sigma ^2}}} \times {(1 + \frac{{n{p_1}{g_1}}}{{n{p_2}{g_3} + {\sigma ^2}}})^{\frac{B}{{RM\ln 2}}}}
\end{aligned},
\end{equation}
\begin{equation} \label{Eq:eq4}
\begin{aligned}
{T_2} &= \frac{{n{p_2}{g_2}}}{{n{p_1}{g_4} + {\sigma ^2}}} \times {e^{\frac{{\frac{B}{R}\log (1 + \frac{{n{p_2}{g_2}}}{{n{p_1}{g_4} + {\sigma ^2}}})}}{M}}}\\
 &= \frac{{n{p_2}{g_2}}}{{n{p_1}{g_4} + {\sigma ^2}}} \times {(1 + \frac{{n{p_2}{g_2}}}{{n{p_1}{g_4} + {\sigma ^2}}})^{\frac{B}{{RM\ln 2}}}}
\end{aligned},
\end{equation}

In addition, to ensure proportion fairness and simplify the model, let
${T _{i \min}=0},\forall i \in \left\{ {1,2, \cdots ,n} \right\}$. Then the utility function can be expressed as follows:

\begin{equation} \label{Eq:eq7}
\begin{aligned}
{U_1} &= {(\frac{{n{p_1}{g_1}}}{{n{p_2}{g_3} + {\sigma ^2}}} \times {(1 + \frac{{n{p_1}{g_1}}}{{n{p_2}{g_3} + {\sigma ^2}}})^{\frac{B}{{RM\ln 2}}}})^{{w_1}}}\\
 &\times {(\frac{{n{p_2}{g_2}}}{{n{p_1}{g_4} + {\sigma ^2}}} \times {(1 + \frac{{n{p_2}{g_2}}}{{n{p_1}{g_4} + {\sigma ^2}}})^{\frac{B}{{RM\ln 2}}}})^{{w_2}}}
\end{aligned}.
\end{equation}

In the first quadrant, the result of multiplying multiple increasing functions is also an increasing function. Otherwise, as for $f = x{(1 + x)^a}{\rm{ }}x > 0,{\rm{ }}a > 0$, we can learn that $\frac{{\partial f}}{{\partial x}} = {(1 + x)^a} + ax{(1 + x)^{a - 1}} > 0$, so the utility function grows with x, $f(x)$ is a monotonically increasing function.
To simplify the system model, we change formula ${U_1}$ into ${F_1}$, which means that the results obtained in our former work \cite{S15} can also be applied to the model considering transmission delay.
\begin{equation} \label{Eq:eq7}
{F_1} = {(\frac{{n{p_1}{g_1}}}{{n{p_2}{g_3} + {\sigma ^2}}})^{{w_1}}}{(\frac{{n{p_2}{g_2}}}{{n{p_1}{g_4} + {\sigma ^2}}})^{{w_2}}}.
\end{equation}

\emph{Theorem 1}.  If there are $n$ BSs, when (28) and (31) are satisfied, that is $(1 + \sqrt 2 )$ times the interference power of the user received from BSs is less than the system noise power, then (15) has the maximum value.

$Proof$: Define ${F_2}={\rm{log}}({F_1})$, which is
\begin{equation} \label{Eq:eq8}
{F_2} = \log ({F_1}) = {w_1}\log (\frac{{n{p_1}{g_1}}}{{n{p_2}{g_3} + {\sigma ^2}}}) + {w_2}\log (\frac{{n{p_2}{g_2}}}{{n{p_1}{g_4} + {\sigma ^2}}}).
\end{equation}

To maximize (15), we should prove (16) has the maximum value firstly.

\begin{equation} \label{Eq:eq8}
\frac{{\partial {F_2}}}{{\partial {p_1}}} = \frac{{{w_1}}}{{{p_1}}} - \frac{{n{g_4}{w_2}}}{{n{p_1}{g_4} + {\sigma ^2}}},
\end{equation}
\begin{equation} \label{Eq:eq8}
\frac{{\partial {F_2}}}{{\partial {p_2}}} = \frac{{{w_2}}}{{{p_2}}} - \frac{{n{g_3}{w_1}}}{{n{p_2}{g_3} + {\sigma ^2}}},
\end{equation}
\begin{equation} \label{Eq:eq8}
A = \frac{{{\partial ^2}{F_2}}}{{\partial p_1^2}} =  - \frac{{{w_1}}}{{p_1^2}} + \frac{{{n^2}{g_4}^2{w_2}}}{{{{(n{p_1}{g_4} + {\sigma ^2})}^2}}},
\end{equation}
\begin{equation} \label{Eq:eq8}
B = 0,
\end{equation}
\begin{equation} \label{Eq:eq8}
C = \frac{{{\partial ^2}{F_2}}}{{\partial p_2^2}} = \frac{{{w_1}{n^2}{g_3}^2}}{{{{(n{p_2}{g_3} + {\sigma ^2})}^2}}} - \frac{{{w_2}}}{{p_2^2}}.
\end{equation}

Above all, we should satisfy:
\begin{equation} \label{Eq:eq8}
AC - {B^2} > 0,
\end{equation}
\begin{equation} \label{Eq:eq8}
A < 0.
\end{equation}
By solving (22)-(23), we can see that,
\begin{equation} \label{Eq:eq8}
ng_4^2({w_2} - {w_1})p_1^2 - 2n{w_1}{p_1}{g_4}{\sigma ^2} - {w_1}{\sigma ^4} < 0,
\end{equation}
\begin{multline} \label{Eq:eq8}
[{n^2}g_4^2({w_2} - {w_1})p_1^2 - 2n{w_1}{g_4}{\sigma ^2}{p_1} - {\sigma ^4}{w_1}]\\
*[{n^2}g_3^2({w_1} - {w_2})p_2^2 - 2n{w_2}{g_3}{\sigma ^2}{p_2} - {\sigma ^4}{w_2}] > 0.
\end{multline}

When all users in the cluster 1 and cluster 2 are high-type (or low-type), ${w_1} = {w_2} = {w_H}({w_L})$, (24) can be simplified as:
\begin{equation} \label{Eq:eq16}
- 2n{g_4}{\sigma ^2}{w_1}{p_1} - {\sigma ^4}{w_1} < 0.
\end{equation}

By simplifying (25), we can see that:
\begin{equation} \label{Eq:eq17}
( - 2n{w_1}{g_4}{\sigma ^2}{p_1} - {\sigma ^4}{w_1})( - 2n{w_2}{g_3}{\sigma ^2}{p_2} - {\sigma ^4}{w_2}) > 0.
\end{equation}

It is obvious that (26) and (27) are true. In conclusion, when the users are all high-type (or low-type), (26) and (27) always satisfy (22) and (23).

When $w{}_1 = {w_2}$, (17) and (18) are always greater than zero. Transmission power allocation takes the upper bound, the utility function reaches the maximum value.

When the user in the cluster 1 is high-type, and the user in the cluster 2 is low-type, i.e., ${w_1} = {w_H}=2$, ${w_2} = {w_L}=1$. Under these circumstances, we should satisfy:
\begin{equation} \label{Eq:eq18}
n{g_3}{p_2} < (1 + \sqrt 2 ){\sigma ^2}.
\end{equation}

To maximize the utility function, that is
\begin{equation} \label{Eq:eq19}
{p_1} = \frac{{{w_1}{\sigma ^2}}}{{({w_2} - {w_1})n{g_4}}},
\end{equation}
\begin{equation} \label{Eq:eq19}
{p_2} = \frac{{{w_2}{\sigma ^2}}}{{n({w_1} - {w_2}){g_3}}}.
\end{equation}

Since the value of ${{w_2} - {w_1}}$ is negative, (29) is negative. When ${p_2} = \frac{{{\sigma ^2}}}{{n{g_3}}}$ and ${p_1}$ takes the upper bound, the utility function reaches the maximum value.

When the user in the cluster 1 is low-type, and the user in the cluster 2 is high-type, ${w_1} = {w_L}=1$, ${w_2} = {w_H}=2$. In this case, we should satisfy:
\begin{equation} \label{Eq:eq20}
n{g_4}{p_1} < (1 + \sqrt 2 ){\sigma ^2}.
\end{equation}

Since the value of ${{w_1} - {w_2}}$ is negative, (30) is negative. When ${p_1} = \frac{{{\sigma ^2}}}{{n{g_4}}}$ and ${p_2}$ takes upper bound, the utility function gets maximum value.

\subsection{Multi-User Multi-BS Case Power Control under NBS}
In this subsection, we consider that there are $M$ MSs and $N$ BSs in a cluster. Under the NBS model, the utility function is\\

\begin{equation} \label{Eq:eq20}
\begin{array}{l}
U = \prod\limits_{i = 1}^M {{{({T_i} - {T_{i\min }})}^{{w_i}}}} \\
s.t.{P_{b,j}} > 0(b = 1,2,...,R;j = 1,2,...,N)\\
\ \ \ \ {\rm{     }}{P_{b,j}} \le {P_{\max }}\\
\ \ \ \ {\rm{     }}{P_{b,j}}{g_{b,j,k}} > {p_0}(b = 1,2,...,R;j = 1,2,...,N;\\
\ \ \ \ {\rm{                           }}k = 1,2,...)\\
\ \ \ \ {\rm{     }}{T_i} > {T_{i\min }}
\end{array}
\end{equation}

where the utility function of user $i$ can be expressed as follows:
\begin{equation} \label{Eq:eq21}
\begin{aligned}
{T_i} &= \frac{{\sum\limits_{j \in CB{S_{b,k}}} {{P_{b,j}}{g_{b,j,k}}} }}{{\sum\limits_{j' \notin CB{S_{b,k}}} {{P_{b,j'}}{g_{b,j',k}}}  + {\sigma ^2}}}\\
&\times {{\rm{e}}^{{\raise0.7ex\hbox{${\frac{B}{R}{\rm{log}}(1 + \frac{{\sum\limits_{j \in CB{S_{b,k}}} {{P_{b,j}}{g_{b,j,k}}} }}{{\sum\limits_{j' \notin CB{S_{b,k}}} {{P_{b,j'}}{g_{b,j',k}}}  + {\sigma ^2}}})}$} \!\mathord{\left/
{\vphantom {{\frac{B}{R}{\rm{log}}(1 + \frac{{\sum\limits_{j \in CB{S_{b,k}}} {{P_{b,j}}{g_{b,j,k}}} }}{{\sum\limits_{j' \notin CB{S_{b,k}}} {{P_{b,j'}}{g_{b,j',k}}}  + {\sigma ^2}}})} M}}\right.\kern-\nulldelimiterspace}
\!\lower0.7ex\hbox{$M$}}}}\\
&= \frac{{\sum\limits_{j \in CB{S_{b,k}}} {{P_{b,j}}{g_{b,j,k}}} }}{{\sum\limits_{j' \notin CB{S_{b,k}}} {{P_{b,j'}}{g_{b,j',k}}}  + {\sigma ^2}}}\\
&\times {(1 + \frac{{\sum\limits_{j \in CB{S_{b,k}}} {{P_{b,j}}{g_{b,j,k}}} }}{{\sum\limits_{j' \notin CB{S_{b,k}}} {{P_{b,j'}}{g_{b,j',k}}}  + {\sigma ^2}}})^{\frac{B}{{RM\ln 2}}}}
\end{aligned}.
\end{equation}

Under this scenario, we define the utility function of NBS as follows:
\begin{equation} \label{Eq:eq22}
\begin{aligned}
U = \prod\limits_{i = 1}^M \begin{array}{l}
(\frac{{\sum\limits_{j \in CB{S_{b,k}}} {{P_{b,j}}{g_{b,j,k}}} }}{{\sum\limits_{j' \notin CB{S_{b,k}}} {{P_{b,j'}}{g_{b,j',k}}}  + {\sigma ^2}}}\\
 \times {(1 + \frac{{\sum\limits_{j \in CB{S_{b,k}}} {{P_{b,j}}{g_{b,j,k}}} }}{{\sum\limits_{j' \notin CB{S_{b,k}}} {{P_{b,j'}}{g_{b,j',k}}}  + {\sigma ^2}}})^{\frac{B}{{RM\ln 2}}}}{)^{{w_i}}}
\end{array}
\end{aligned}.
\end{equation}

Similarly, the result will be an increasing function in the first quadrant. To simplify the system model, we consider that ${T _{\min 1}}=0$. Therefore, ${U}$ can be changed into ${F_1}$.
\begin{equation} \label{Eq:eq80}
{F_1} = {\prod\limits_{i = 1}^M {(\frac{{\sum\limits_{j \in CB{S_{b,k}}} {{P_{b,j}}{g_{b,j,k}}} }}{{\sum\limits_{j' \notin CB{S_{b,k}}} {{P_{b,j'}}{g_{b,j',k}}}  + {\sigma ^2}}})} ^{{w_i}}}.
\end{equation}

Define ${F_2}={\rm{log}}({F_1})$, which is
\begin{equation} \label{Eq:eq80}
{F_2} = \log ({F_1}) = \sum\limits_{i = 1}^M {{w_i}\log (\frac{{\sum\limits_{j \in CB{S_{b,k}}} {{P_{b,j}}{g_{b,j,k}}} }}{{\sum\limits_{j^\prime \notin CB{S_{b,k}}} {{P_{b,j^\prime }}{g_{b,j^\prime ,k}}}  + {\sigma ^2}}})}.
\end{equation}

To maximize (34), we need to prove (36) has the maximum value firstly.

\begin{equation} \label{Eq:eq23}
\frac{{\partial {F_2}}}{{\partial {P_{b,j}}}} = \sum\limits_{i = 1}^M {\frac{{{w_i}}}{{{P_{b,j}}}}}  > 0,
\end{equation}

we have
\begin{equation} \label{Eq:eq24}
\frac{{{\partial ^2}{F_2}}}{{\partial P_{b,j}^2}} =  - \sum\limits_{i = 1}^M {\frac{{{w_i}}}{{{P_{b,j}}^2}}}  < 0.
\end{equation}

From (37) and (38), we can see that the utility function is a monotonically increasing function, although the growth rate of this utility function drops. Hence, the transmitting power should be the maximum power of the BSs under the multi-BSs multi-users scenario.

\section{Numerical Results}

\subsection{Simulation Setup}

Simulations were done using MATLAB 2010. By comparing the performance of the proposed scheme, the throughput and transmission delay are both improved significantly.
To manifest the effectiveness of the algorithm proposed in the ultra-dense networks, we compare the throughput of edge users at different distances.
The simulation parameters are detailedly described as follows: The CoMP system consists of 19 cells, a frequency reuse factor of 1 \cite{S22}, and the cell radius ranges from 0.05 km to 0.5 km. There are 100 users randomly distributed in each cell. In this paper, we set the system bandwidth as $B=3$MHz and the pathloss as $148.1 + 37.6{\log _{10}}d$, where $d$ is the distance from BS to UE.
We calculate the throughput and transmission delay assuming the size of video is 100MB.
\begin{table}[t]\label{Tab1}
\centering
\caption{Simulation parameters}
\begin{tabular}{c|c}
\hline
Parameter & Value\\
\hline
Number of cells  & 19 \\
\hline
System bandwidth & 3MHz\\
\hline
BS maximum power & 43dBm\\
\hline
Number of PRBs  & 15\\
\hline
Number of users per cell  & 100\\
\hline
Cell radius & 0.05Km-0.5Km \\
\hline
Path loss model  & $148.1 + 37.6{\log _{10}}d$ \\
\hline
Frequency reuse factor  & 1 \\
\hline
The size of the video file  & 100MB \\
\hline
Link  & Down link \\
\hline
Antenna gain & 5dB\\
\hline
Noise power spectral density & -174dBm/Hz\\
\hline
\end{tabular}
\end{table}
\subsection{Simulation Results}
The edge users are served by several neighboring BSs separately, as shown in Fig. 6. To indicate the effect of the algorithm exploiting AP cluster, we consider that there is just one BS per cell which is at the center of the cell. The green point shown in Fig. 6 represents the BS in each cell. The points that the lines cross are the center of clusters, which represent the edge users of the cells are served. Different kinds of colours stand for different clusters.

\begin{figure}[t]
\centering
\includegraphics[width=0.5 \textwidth]{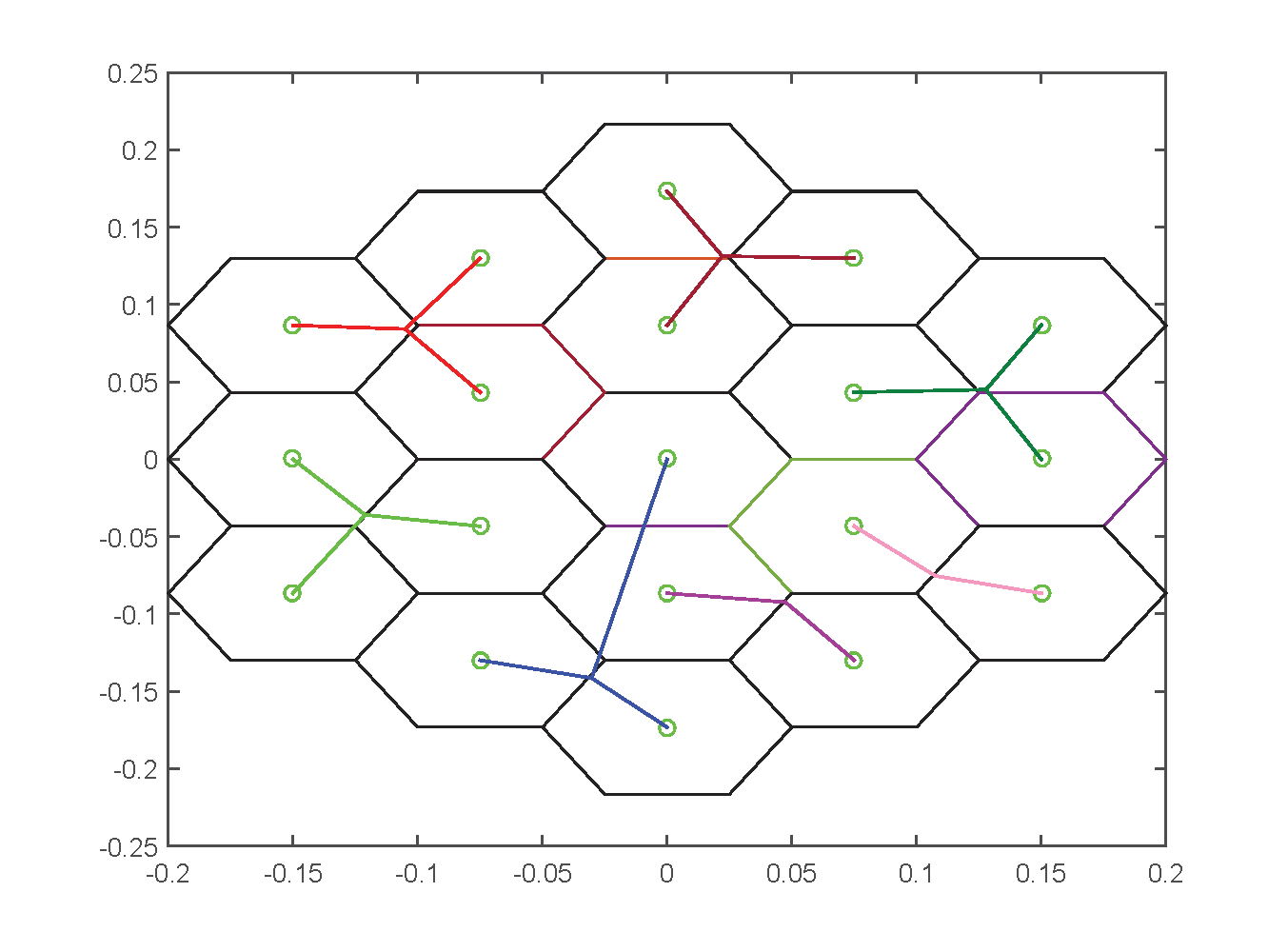}
\caption{Result of cluster using AP cluster algorithm for SA} \label{Fig:inorderto50m}
\end{figure}

\begin{figure}[t]
\centering
\includegraphics[width=0.5 \textwidth]{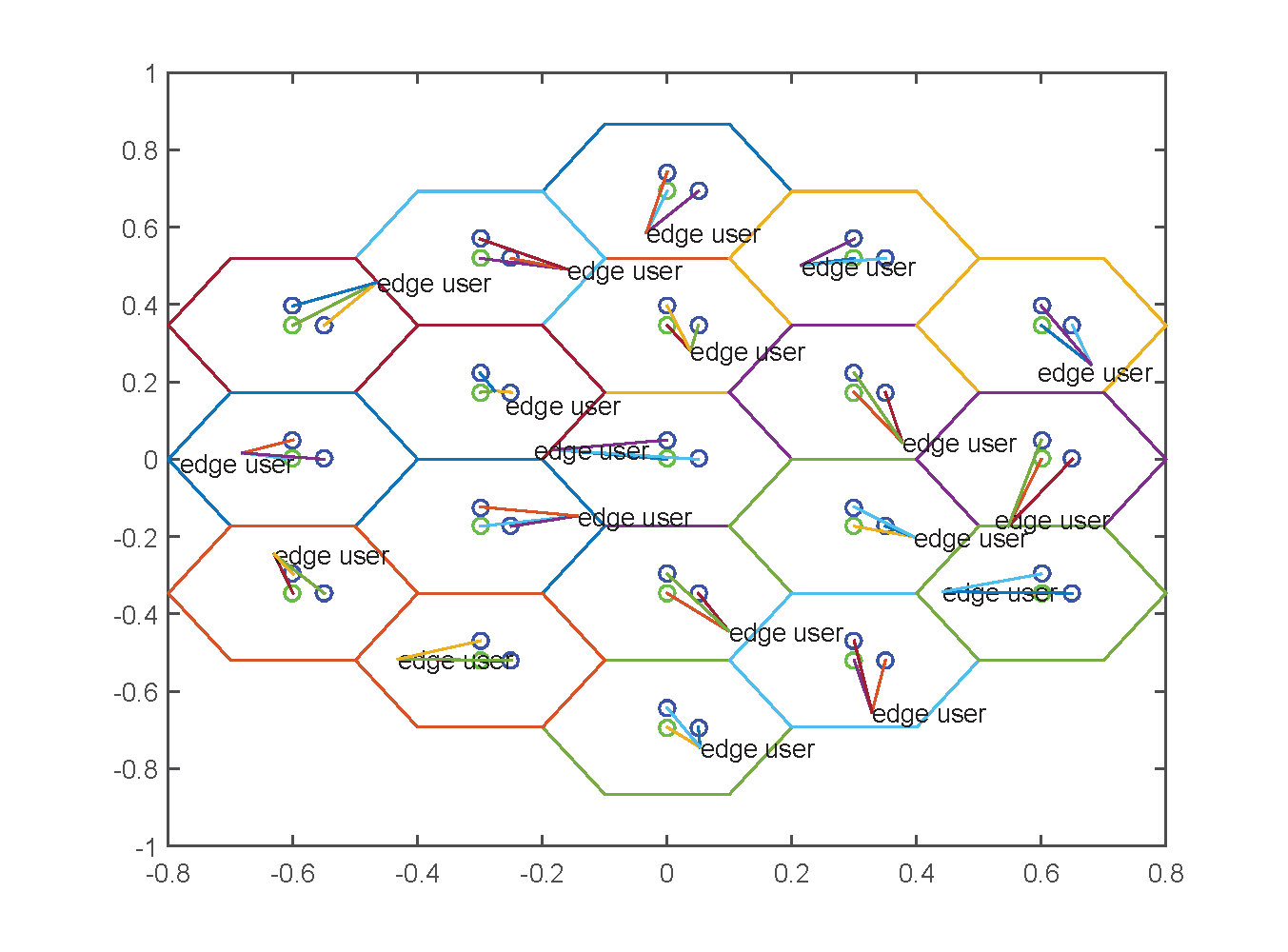}
\caption{Result of cluster using AP cluster algorithm for NSA} \label{Fig:inorderto50m}
\end{figure}

Due to the advantages of SBS, 5G introduces SBS into the ultra-dense networks. Fig. 7 shows the scenario where the green point at the center of the cell represents MBS and the neighbouring blue points around the MBS represent SBSs. The distance between the MBS and the SBS in the same cell is 50m and the cell radius is 200m. The centers of the clusters represent edge users. The lines connected to the center express the service relationship. As can be seen in Fig.6 and Fig.7, the sizes of clusters obtained by the AP clustering algorithm are different.

\begin{figure}[t]
\centering
\includegraphics[width=0.55 \textwidth]{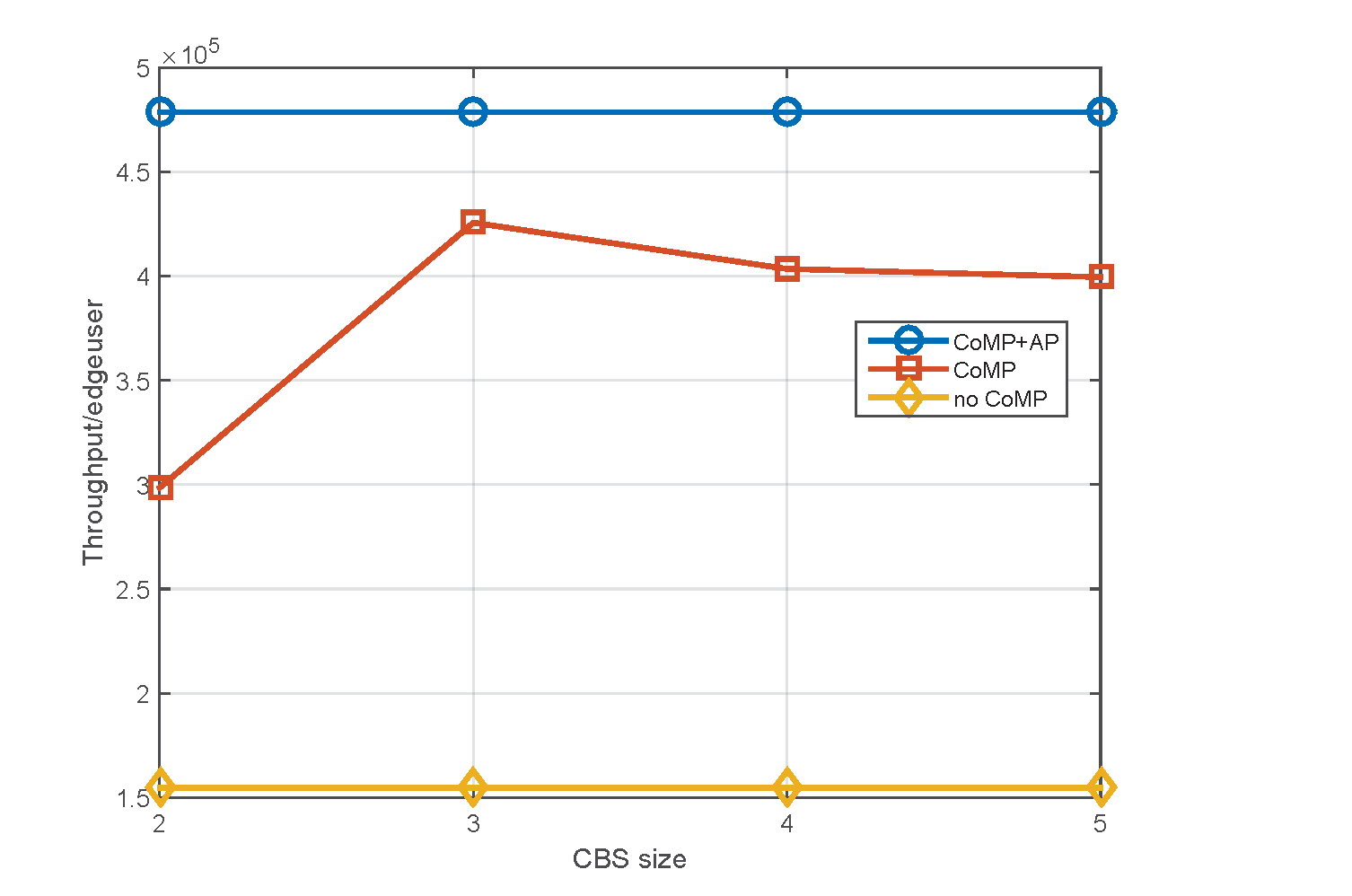}
\caption{Throughput comparison for SA} \label{Fig:compare0513}
\end{figure}

The interference becomes more severe due to the denser networks. To compare the average rate of edge users under these algorithms, we simulate the scenario of 19 cells, and the cell radius is 50m.
It can be observed from Fig.8 that the throughput changes according to the size of clusters. The line which expresses performance of the algorithm with CoMP tends to be gentle when the cluster size becomes 3.
The average throughput of the edge users using AP cluster algorithm in CoMP is higher than the highest throughput of the edge users using algorithm without AP cluster in CoMP.

\begin{figure}[t]
\centering
\includegraphics[width=0.6 \textwidth]{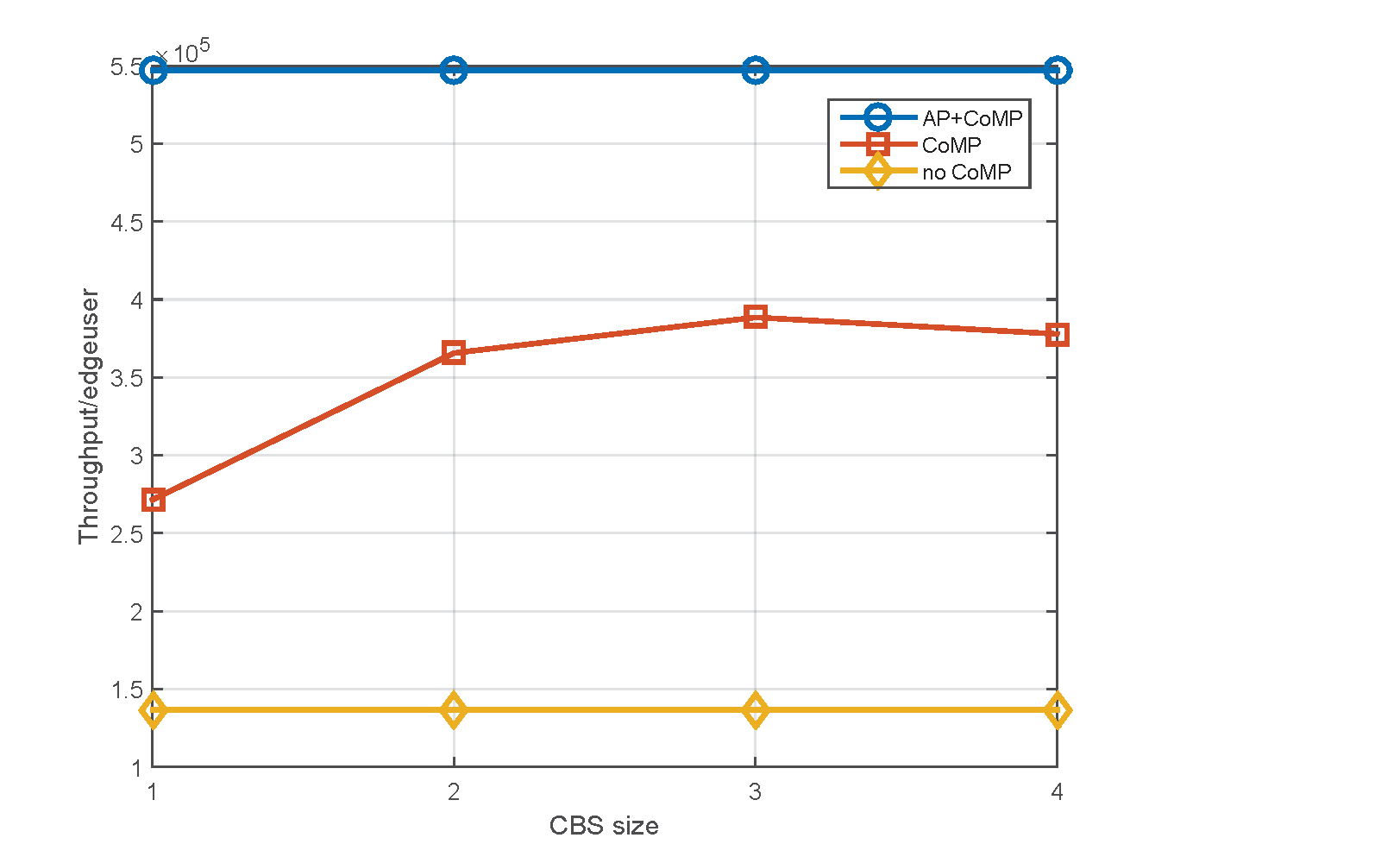}
\caption{Throughput comparison for NSA} \label{Fig:compare0513}
\end{figure}

From the numerical results presented in Fig. 9, we can conclude that the algorithm exploiting AP cluster and CoMP can increase the throughput of edge users drastically. The simulation scenario is the same as that of Fig. 7. The performance gain of the algorithm exploiting AP cluster for 5G NSA is better than the performance gain in the scenario for 5G SA. For example, when the CBS size is 4 in common CoMP, the average throughput of edge user is almost $4*10^5$ bps which uses common transmission schedule scheme. The average throughput of edge user is $5.5*10^5$ bps which uses the scheme we proposed, while it is only $1.4*10^5$ bps of the edge user without CoMP. It can be concluded that the AP cluster algorithm can adapt to the complex scenario dynamically.

\begin{figure}[t]
\centering
\includegraphics[width=0.52 \textwidth]{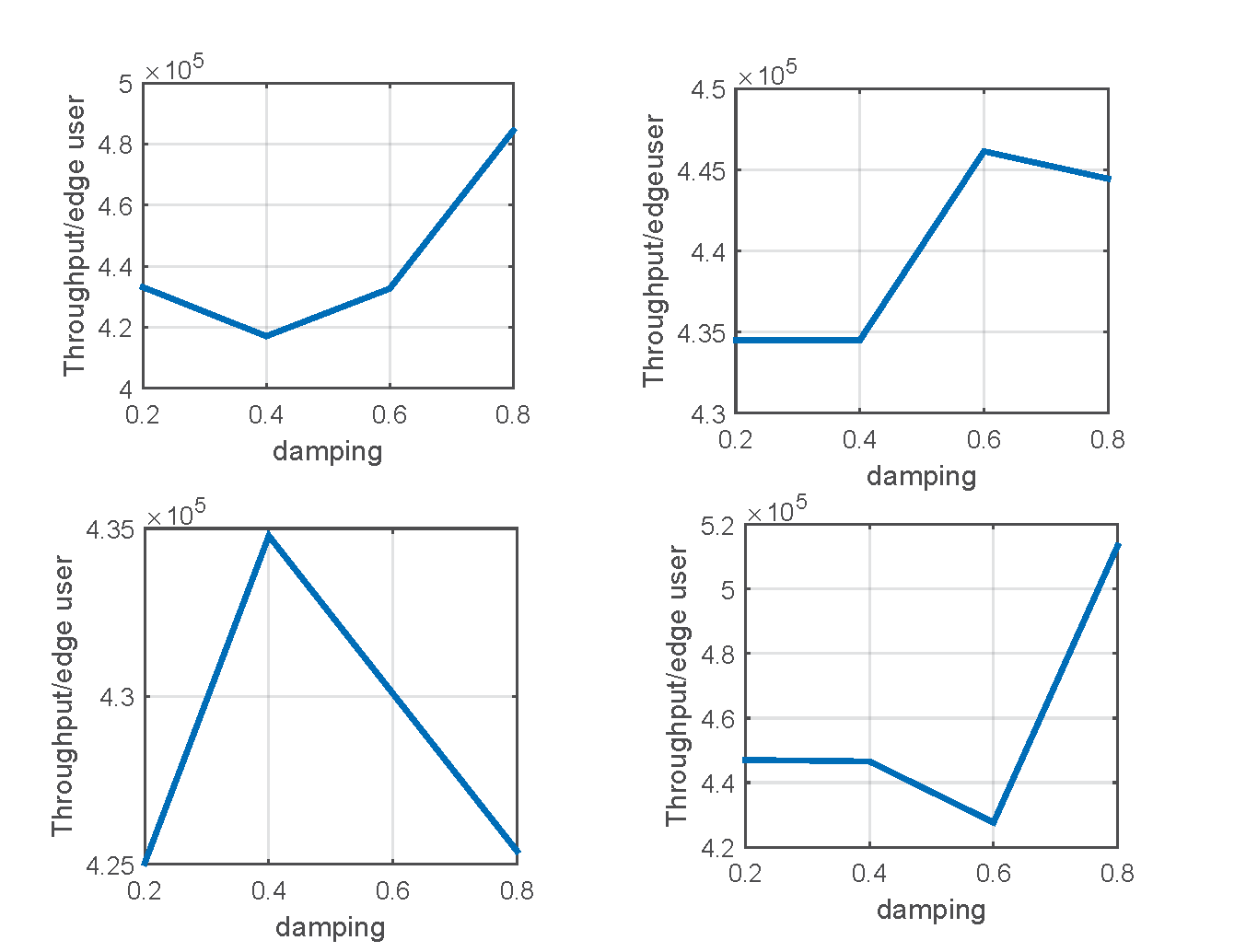}
\caption{The impact of damping} \label{Fig:fig6}
\end{figure}

Fig. 10 shows the impact of damping factor. But the lines do not have the same trend. To reduce the impact of random errors, we simulate via cycling experiment. Hence, we consider that the damping cannot impact the throughput of edge users.

\begin{figure}[!ht]
\centering
\includegraphics[width=0.45 \textwidth]{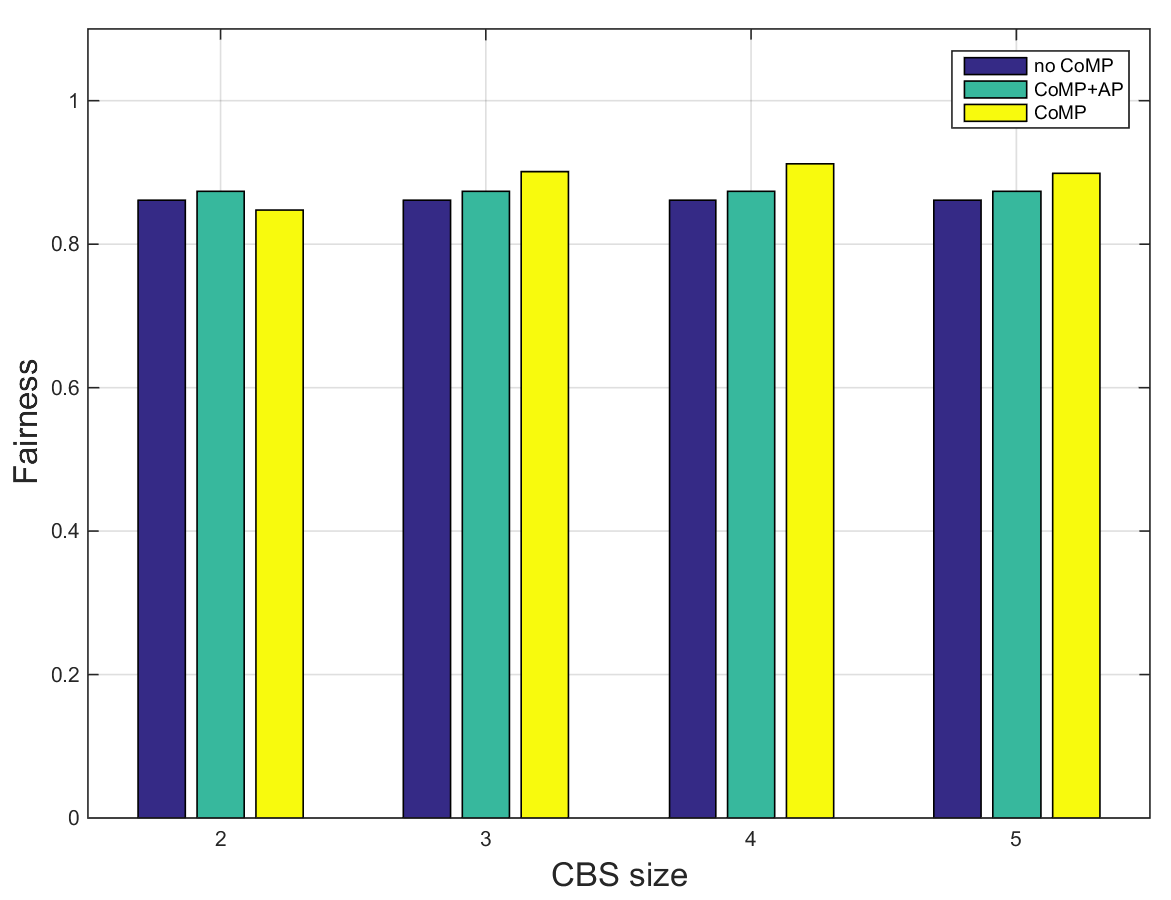}
\caption{Fairness comparison} \label{Fig:fig6}
\end{figure}
\begin{figure}[t]
\centering
\includegraphics[width=0.6 \textwidth]{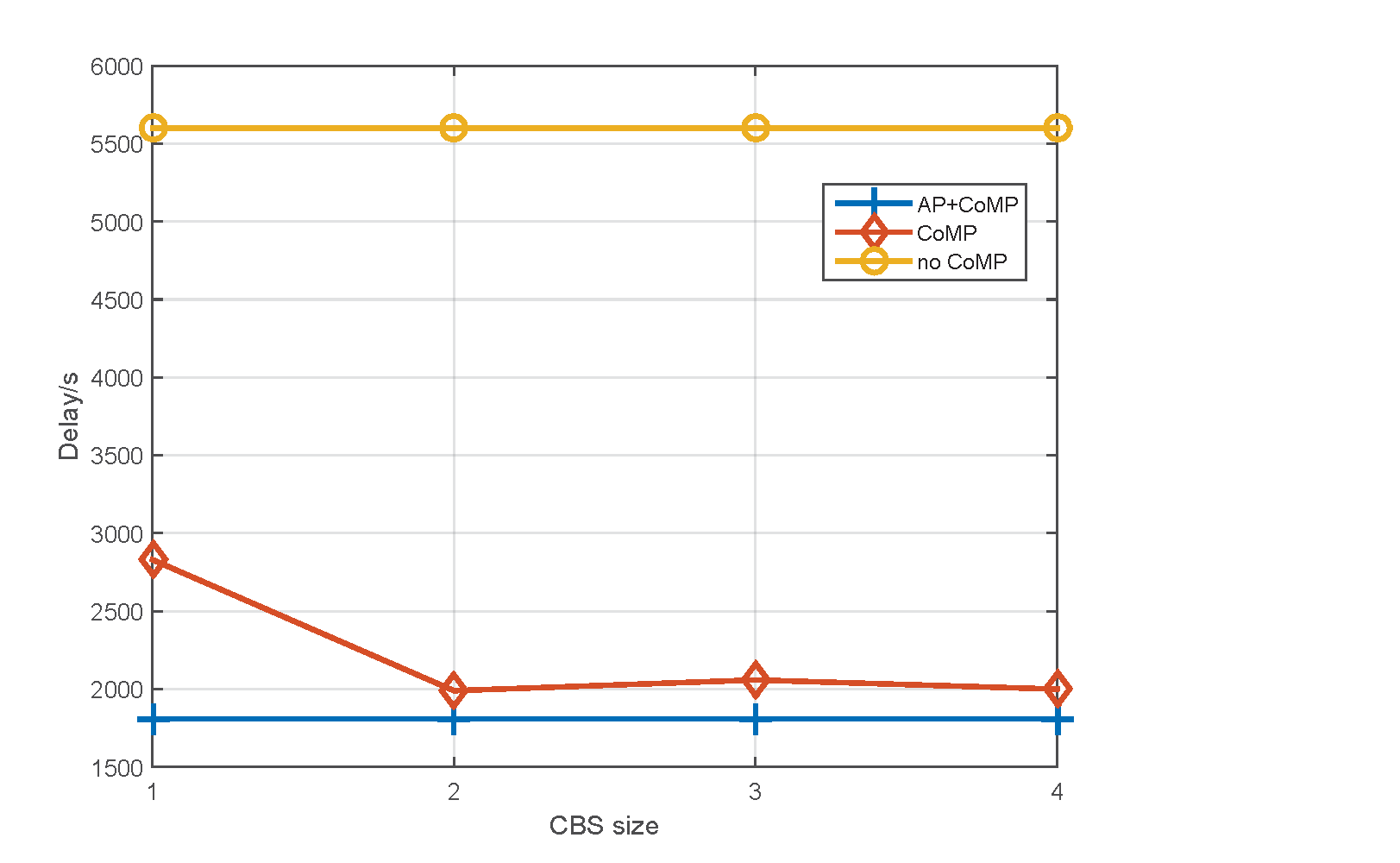}
\caption{Transmission delay comparison for SA} \label{Fig:delay}
\end{figure}
\begin{figure}[t]
\centering
\includegraphics[width=0.6 \textwidth]{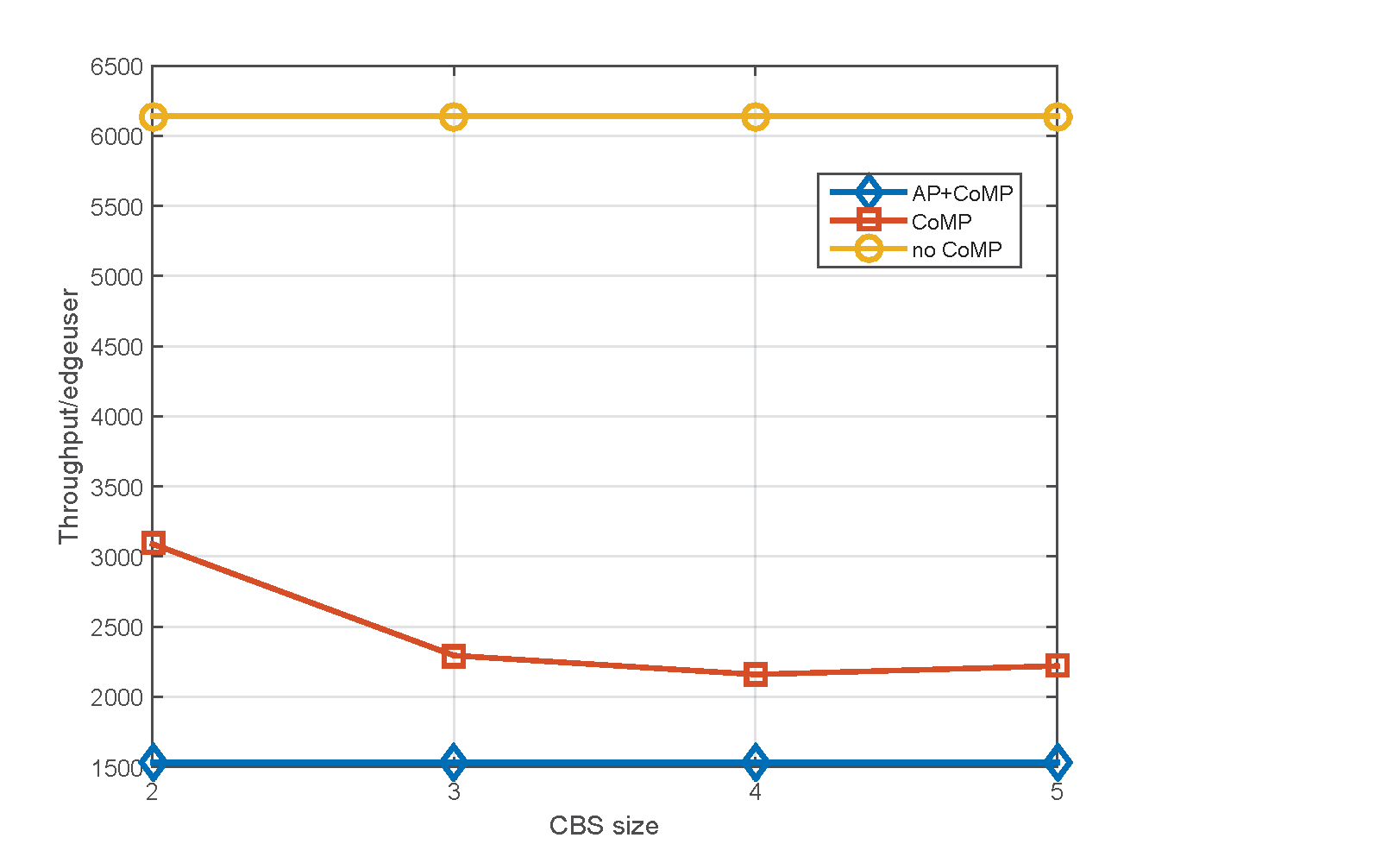}
\caption{Transmission delay comparison for NSA} \label{Fig:delay}
\end{figure}
\begin{figure}[!ht]
\centering
\includegraphics[width=0.52 \textwidth]{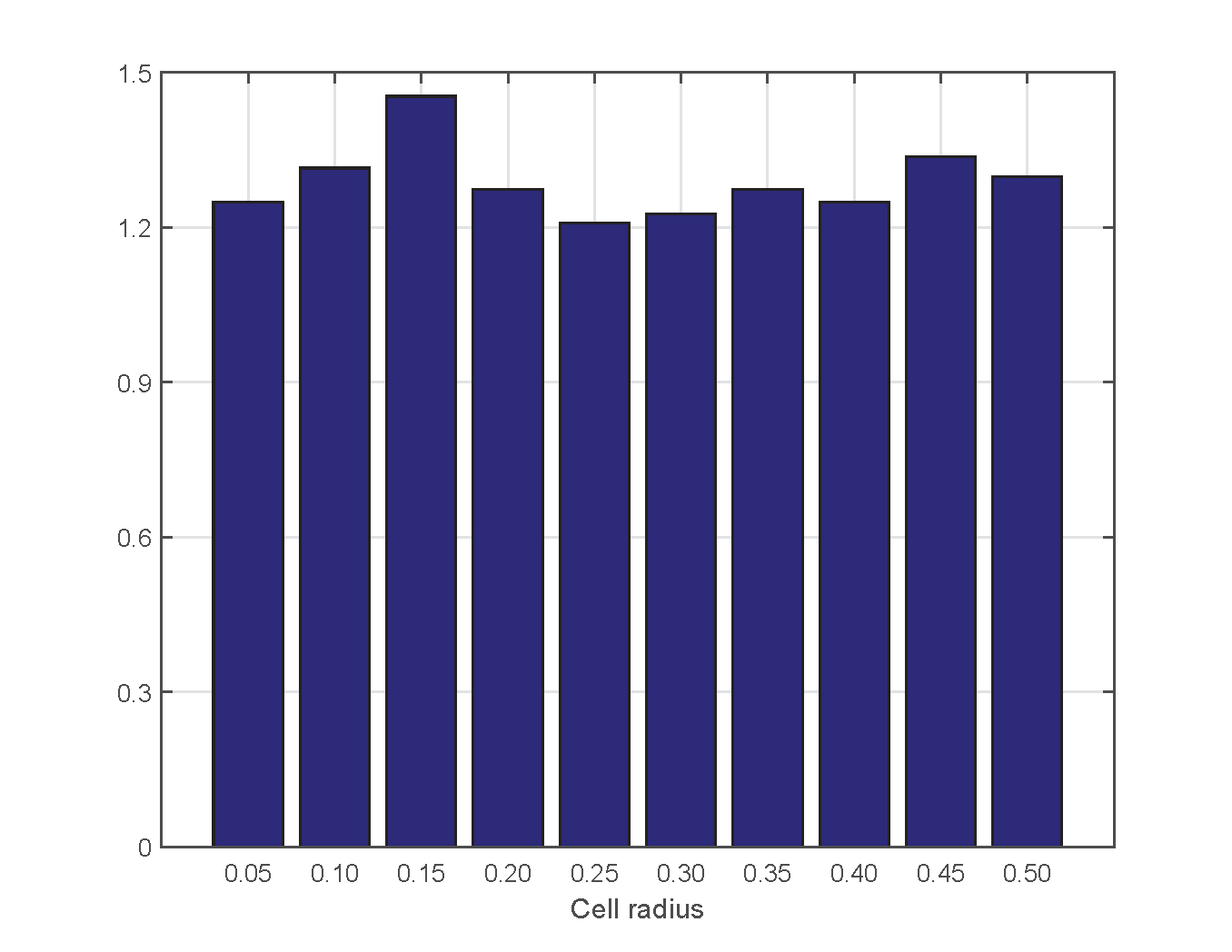}
\caption{Performance gain of algorithm with AP over common CoMP} \label{Fig:CoMP_over_no_CoMP_and_AP_over_CoMP}
\end{figure}

To compare the fairness among these algorithms. Let the average users' transmission rates be ${\bar R_1},...,{\bar R_K}$. Define the Jain's fairness index as $\frac{{{{(\sum\nolimits_{k = 1}^K {{{\bar R}_k}} )}^2}}}{{K\sum\nolimits_{k = 1}^K {\bar R_k^2} }}$ \cite{S23}.
NBS can improve the fairness of users, which is achieved by power control. However, the simulation results shown in Fig. 11 indicate that the NBS fairness is not obviously higher than the users using other algorithms, which can be attributed to the similarity of transmission power of BSs between the common CoMP algorithm and the algorithm under multi-user multi-BS conditions.

Note that the transmission delay can be obtained via dividing the size of a video file by the transmission rate of the edge users. To compare the delay of these algorithms, we set the size of a video file as 100MB and simulate the transmission scenario. It is evident from Fig. 12 that the algorithm we proposed has less transmission delay than others because it can observably improve the throughput of edge users.

The transmission delay in 5G NSA deployment is shown in Fig. 13, we can see that our algorithm has the lowest transmission delay among these algorithms. Because the algorithm we proposed can increase the transmission rate of edge users.

With the development of the network, the distance between BSs becomes smaller, the interference among edge users gets stronger, and the analysis becomes more complex. We simulate the performance of the average throughput of edge users in 5G NSA deployment. As shown in Fig. 14, we can find that the performance of the algorithm with AP cluster is 1.2 times more than the algorithm with common CoMP regardless of the change of cell radius. In addition, Fig. 14 also proves the algorithm is effective in the ultra-dense networks.

\section{Conclusion}

In this paper, we proposed a user-centric dynamic framework including downlink transmission scheduling and power control for improving the performance of edge users in 5G CoMP systems.
Specifically, a distributed dynamic transmission algorithm was firstly proposed.
Based on this, a transmission scheduling scheme took AP cluster algorithm into consideration. And we utilized NBS to deduce the power control scheme to increase the fairness of users.
We built the utility function of the system model based on users' throughput and transmission delay and proved the existence of the unique Nash equilibrium solution of this function.
Finally, the simulation results showed the proposed approach can dramatically improve the performance of edge users.

\end{document}